\documentclass[aps,preprint]{revtex4}
\usepackage{amssymb,epsf}
\usepackage{amsmath}
\usepackage{graphicx}

\begin{document}
\title{Topological Black Holes in Lovelock-Born-Infeld Gravity}
\author{M. H. Dehghani$^{1,2}$\footnote{email address:
mhd@shirazu.ac.ir}, N. Alinejadi$^{1}$ and S. H. Hendi$^{3}$\footnote{email address:
hendi@mail.yu.ac.ir}}

\affiliation{$^1$Physics Department and Biruni Observatory,
College of Sciences, Shiraz
University, Shiraz 71454, Iran \\
$^2$Research Institute for Astrophysics and Astronomy of Maragha
(RIAAM), Maragha, Iran\\
$^3$ Physics Department, College of Sciences, Yasouj University, Yasouj 75914, Iran}
\begin{abstract}
In this paper, we present topological black holes of
third order Lovelock gravity in the presence of cosmological constant and
nonlinear electromagnetic Born-Infeld field. Depending on the metric parameters, these solutions
may be interpreted as black hole solutions with inner and outer event
horizons, an extreme black hole or naked singularity. We investigate
the thermodynamics of asymptotically flat solutions and show that
the thermodynamic and conserved quantities of these black holes
satisfy the first law of thermodynamic. We also endow the Ricci flat solutions
with a global rotation and calculate the finite action and conserved quantities
of these class of solutions by using the counterterm method. We compute the
entropy through the use of the Gibbs-Duhem relation and find that the entropy
obeys the area law. We obtain a Smarr-type formula for the mass as a
function of the entropy, the angular momenta, and the charge, and compute
temperature, angular velocities, and electric potential and show that these
thermodynamic quantities coincide with their values which are computed
through the use of geometry. Finally, we perform a stability analysis for
this class of solutions in both the canonical and the grand-canonical ensemble and show that the
presence of a nonlinear electromagnetic field and higher curvature terms has
no effect on the stability of the black branes, and they are stable in the
whole phase space.
\end{abstract}

\maketitle
\section{Introduction}

Over the last few years, several extra-dimensional models have been
introduced in an attempt to deal with the hierarchy problem. These models can
lead to rather unique and spectacular signatures at Terascale colliders such
as the LHC and ILC. In higher dimensions, it is known that the
Einstein-Hilbert (EH) Lagrangian, $R$, can only be regarded as the first order
term in an effective action, so one may on general grounds expect that as
one probes energies approaching the fundamental scale, significant
deviations from EH expectations are likely to appear. This motivates one to
consider the more general class of gravitational action:
\[
I_{G}=\frac{1}{16\pi }\int d^{n+1}x\sqrt{-g} \mathcal{F}(R,R_{\mu \nu
}R^{\mu \nu },R_{\mu \nu \rho \sigma }R^{\mu \nu \rho \sigma }).\]
The presence of higher curvature terms can also be seen in the
renormalization of quantum field theory in curved spacetime \cite{Birrel},
or in the construction of low energy effective action of string theory \cite
{Green}. Among the higher curvature gravity theories, the so-called Lovelock
gravity is quite special, whose Lagrangian consist of the dimensionally
extended Euler densities. This Lagrangian is obtained by Lovelock as he
tried to calculate the most general tensor that satisfies properties of
Einstein's tensor in higher dimensions \cite{Lovelock}. Since the Lovelock
tensor contains derivatives of metrics of order not higher than two, the
quantization of linearized Lovelock theory is free of ghosts \cite{Boulware}%
. Thus, it is natural to study the effects of higher curvature terms on the
properties and thermodynamics of black holes.

Accepting the nonlinear terms of the invariants constructed by Riemann
tensor on the gravity side of the action, it seems natural to add the
nonlinear terms in the matter action too. Thus, in the presence of
an electromagnetic field, it is worthwhile to apply the action of Born-Infeld \cite{BI}
instead of the Maxwell action. In this paper, we generalize static and rotating
black hole solutions of third order Lovelock gravity in the
presence of Maxwell field \cite{DSh,DM1} to the case of these
solutions in the presence of nonlinear electromagnetic fields. Indeed, it is interesting to
explore new black hole solutions in higher curvature gravity and investigate
which properties of black holes are peculiar to Einstein gravity, and which
are robust features of all generally covariant theories of gravity. The
first aim to relate the nonlinear electrodynamics and gravity has been done
by Hoffmann \cite{Hoffmann}. He obtained a solution of the Einstein
equations for a pointlike Born-Infeld charge, which is devoid of the
divergence of the metric at the origin that characterizes the
Reissner-Nordstr\"{o}m solution. However, a conical singularity remained
there, as it was later objected by Einstein and Rosen. The spherically
symmetric solutions in Einstein-Born-Infeld gravity with or without a
cosmological constant have been considered by many authors \cite{EBI,Dey},
while the rotating solutions of this theory is investigated in \cite
{Rastegar}. Also, these kinds of solutions in the presence of a dilaton field
have been introduced in \cite{DHSR}. The static black hole solutions of
Gauss-Bonnet-Born-Infeld gravity have been constructed in Ref. \cite{Wil2},
and the rotating solution of this theory has been considered in \cite{DH}.

The out line of our paper is as follows. We present the topological black
holes of third order Lovelock gravity in the presence of Born-Infeld field
in Sec. \ref{Topol}. In Sec. \ref{flat}, we calculate the thermodynamic
quantities of asymptotically flat solutions and investigate the first law of
thermodynamics. In Sec. \ref{rotating} we introduce the rotating
solutions with flat horizon and compute the thermodynamic and conserved
quantities of them. We also perform a stability analysis of the solutions
both in canonical and grand canonical ensemble. We finish our paper with
some concluding remarks.

\section{ Topological Black Holes}

\label{Topol}

The action of third order Lovelock gravity in the presence of nonlinear
Born-Infeld electromagnetic field is
\begin{equation}
I_{G}=\frac{1}{16\pi }\int d^{n+1}x\sqrt{-g}\left( -2\Lambda +\mathcal{L}%
_{1}+\alpha _{2}\mathcal{L}_{2}+\alpha _{3}\mathcal{L}_{3}+L(F)\right),
\label{Act1}
\end{equation}
where $\Lambda $ is the cosmological constant, $\alpha _{2}$ and\textbf{\ }$%
\alpha _{3}$\textbf{\ }are the second and third order Lovelock coefficients,
$\mathcal{L}_{1}=R$ is just the Einstein-Hilbert Lagrangian, $\mathcal{L}%
_{2}=R_{\mu \nu \gamma \delta }R^{\mu \nu \gamma \delta }-4R_{\mu \nu
}R^{\mu \nu }+R^{2}$ is the Gauss-Bonnet Lagrangian,
\begin{eqnarray}
\mathcal{L}_{3} &=&2R^{\mu \nu \sigma \kappa }R_{\sigma \kappa \rho \tau }R_{%
\phantom{\rho \tau }{\mu \nu }}^{\rho \tau }+8R_{\phantom{\mu \nu}{\sigma
\rho}}^{\mu \nu }R_{\phantom {\sigma \kappa} {\nu \tau}}^{\sigma \kappa }R_{%
\phantom{\rho \tau}{ \mu \kappa}}^{\rho \tau }+24R^{\mu \nu \sigma \kappa
}R_{\sigma \kappa \nu \rho }R_{\phantom{\rho}{\mu}}^{\rho }  \nonumber \\
&&+3RR^{\mu \nu \sigma \kappa }R_{\sigma \kappa \mu \nu }+24R^{\mu \nu
\sigma \kappa }R_{\sigma \mu }R_{\kappa \nu }+16R^{\mu \nu }R_{\nu \sigma
}R_{\phantom{\sigma}{\mu}}^{\sigma }-12RR^{\mu \nu }R_{\mu \nu }+R^{3}
\label{L3}
\end{eqnarray}
is the third order Lovelock Lagrangian, and $L(F)$ is the Born-Infeld
Lagrangian given as
\begin{equation}
L(F)=4\beta ^{2}\left( 1-\sqrt{1+\frac{F^{2}}{2\beta ^{2}}}\right).
\end{equation}
In the limit $\beta \rightarrow \infty $, $L(F)$ reduces to the standard
Maxwell form $L(F)=-F^{2}$, where $F_{\mu \nu }=\partial _{\mu }A_{\nu
}-\partial _{\nu }A_{\mu }$. Varying the action (\ref{Act1}) with respect to
the metric tensor $g_{\mu \nu }$ and electromagnetic vector field $A_{\mu }$
the equations of gravitation and electromagnetic fields are obtained as:
\begin{equation}
G_{\mu \nu }^{(1)}+\Lambda g_{\mu \nu }+\alpha _{2}G_{\mu \nu }^{(2)}+\alpha
_{3}G_{\mu \nu }^{(3)}=\frac{1}{2}g_{\mu \nu }L(F)+\frac{2F_{\mu \lambda }F_{%
{\nu }}^{\lambda }}{\sqrt{1+\frac{F^{2}}{2\beta ^{2}}}},  \label{Geq}
\end{equation}
\begin{equation}
\partial _{\mu }\left( \frac{\sqrt{-g}F^{\mu \nu }}{\sqrt{1+\frac{F^{2}}{%
2\beta ^{2}}}}\right) =0,  \label{BIeq}
\end{equation}
where $G_{\mu \nu }^{(1)}$ is the Einstein tensor, and $G_{\mu \nu }^{(2)}$
and $G_{\mu \nu }^{(3)}$ are the second and third order Lovelock tensors
given as \cite{Muller}:
\begin{equation}
G_{\mu \nu }^{(2)}=2(R_{\mu \sigma \kappa \tau }R_{\nu }^{\phantom{\nu}%
\sigma \kappa \tau }-2R_{\mu \rho \nu \sigma }R^{\rho \sigma }-2R_{\mu
\sigma }R_{\phantom{\sigma}\nu }^{\sigma }+RR_{\mu \nu })-\frac{1}{2}%
\mathcal{L}_{2}g_{\mu \nu },  \label{Love2}
\end{equation}
\begin{eqnarray}
G_{\mu \nu }^{(3)} &=&-3(4R^{\tau \rho \sigma \kappa }R_{\sigma \kappa
\lambda \rho }R_{\phantom{\lambda }{\nu \tau \mu}}^{\lambda }-8R_{%
\phantom{\tau \rho}{\lambda \sigma}}^{\tau \rho }R_{\phantom{\sigma
\kappa}{\tau \mu}}^{\sigma \kappa }R_{\phantom{\lambda }{\nu \rho \kappa}%
}^{\lambda }+2R_{\nu }^{\phantom{\nu}{\tau \sigma \kappa}}R_{\sigma \kappa
\lambda \rho }R_{\phantom{\lambda \rho}{\tau \mu}}^{\lambda \rho }  \nonumber
\\
&&-R^{\tau \rho \sigma \kappa }R_{\sigma \kappa \tau \rho }R_{\nu \mu }+8R_{%
\phantom{\tau}{\nu \sigma \rho}}^{\tau }R_{\phantom{\sigma \kappa}{\tau \mu}%
}^{\sigma \kappa }R_{\phantom{\rho}\kappa }^{\rho }+8R_{\phantom
{\sigma}{\nu \tau \kappa}}^{\sigma }R_{\phantom {\tau \rho}{\sigma \mu}%
}^{\tau \rho }R_{\phantom{\kappa}{\rho}}^{\kappa }  \nonumber \\
&&+4R_{\nu }^{\phantom{\nu}{\tau \sigma \kappa}}R_{\sigma \kappa \mu \rho
}R_{\phantom{\rho}{\tau}}^{\rho }-4R_{\nu }^{\phantom{\nu}{\tau \sigma
\kappa }}R_{\sigma \kappa \tau \rho }R_{\phantom{\rho}{\mu}}^{\rho
}+4R^{\tau \rho \sigma \kappa }R_{\sigma \kappa \tau \mu }R_{\nu \rho
}+2RR_{\nu }^{\phantom{\nu}{\kappa \tau \rho}}R_{\tau \rho \kappa \mu }
\nonumber \\
&&+8R_{\phantom{\tau}{\nu \mu \rho }}^{\tau }R_{\phantom{\rho}{\sigma}%
}^{\rho }R_{\phantom{\sigma}{\tau}}^{\sigma }-8R_{\phantom{\sigma}{\nu \tau
\rho }}^{\sigma }R_{\phantom{\tau}{\sigma}}^{\tau }R_{\mu }^{\rho }-8R_{%
\phantom{\tau }{\sigma \mu}}^{\tau \rho }R_{\phantom{\sigma}{\tau }}^{\sigma
}R_{\nu \rho }-4RR_{\phantom{\tau}{\nu \mu \rho }}^{\tau }R_{\phantom{\rho}%
\tau }^{\rho }  \nonumber \\
&&+4R^{\tau \rho }R_{\rho \tau }R_{\nu \mu }-8R_{\phantom{\tau}{\nu}}^{\tau
}R_{\tau \rho }R_{\phantom{\rho}{\mu}}^{\rho }+4RR_{\nu \rho }R_{%
\phantom{\rho}{\mu }}^{\rho }-R^{2}R_{\nu \mu })-\frac{1}{2}\mathcal{L}%
_{3}g_{\mu \nu }.  \label{Love3}
\end{eqnarray}

Here we want to obtain the $(n+1)$-dimensional static solutions of Eqs. (\ref
{Geq}) and (\ref{BIeq}). We assume that the metric has the following form:
\begin{equation}
ds^{2}=-f(r)dt^{2}+\frac{dr^{2}}{f(r)}+r^{2}d\Omega ^{2},  \label{met}
\end{equation}
where
\[
d\Omega ^{2}=\left\{
\begin{array}{cc}
d\theta _{1}^{2}+\sum\limits_{i=2}^{n-1}\prod\limits_{j=1}^{i-1}\sin
^{2}\theta _{j}d\theta _{i}^{2} & k=1 \\
d\theta _{1}^{2}+\sinh ^{2}\theta _{1}d\theta _{2}^{2}+\sinh ^{2}\theta
_{1}\sum\limits_{i=3}^{n-1}\prod\limits_{j=2}^{i-1}\sin ^{2}\theta
_{j}d\theta _{i}^{2} & k=-1 \\
\sum\limits_{i=1}^{n-1}d\phi _{i}^{2} & k=0
\end{array}
\right\}\]
represents the line element of an $(n-1)$-dimensional hypersurface
with constant curvature $(n-1)(n-2)k$ and volume $V_{n-1}$.

Using Eq. (\ref{BIeq}), one can show that the vector
potential can be written as
\begin{equation}
A_{\mu }=-\sqrt{\frac{(n-1)}{2n-4}}\frac{q}{r^{n-2}}\digamma (\eta )\delta
_{\mu }^{0},  \label{Amu1}
\end{equation}
where $q$ is an integration constant which is related to the charge
parameter and
\[
\eta =\frac{{(n-1)(n-2)q^{2}}}{2\beta ^{2}r^{2n-2}}.
\]
In Eq. (\ref{Amu1}) and throughout the paper, we use the following abbreviation for
the hypergeometric function
\begin{equation}
{_{2}F_{1}}\left( {%
\left[ \frac{1}{2},\frac{n-2}{2n-2}\right] ,\left[ \frac{3n-4}{2n-2}\right]
,-z }\right) =\digamma (z).  \label{hyp}
\end{equation}
The hypergeometric function $\digamma (\eta ){\rightarrow 1}$ as $\eta
\rightarrow 0$ ($\beta \rightarrow \infty $) and therefore $A_{\mu }$ of Eq.
(\ref{Amu1}) reduces to the gauge potential of Maxwell field. One may show
that the metric function
\begin{equation}
f(r)=k+\frac{r^{2}}{\alpha }\left( 1-g(r)^{1/3}\right),  \label{fr}
\end{equation}
\begin{equation}
g(r)=1+\frac{3\alpha m}{r^{n}}-\frac{12\alpha \beta ^{2}}{n(n-1)}\left[ 1-%
\sqrt{1+\eta }-\frac{\Lambda }{2\beta ^{2}}+\frac{(n-1)\eta }{(n-2)}\digamma
(\eta )\right]   \label{gr}
\end{equation}
satisfies the field equations (\ref{Geq}) in the special case
\begin{eqnarray*}
\alpha _{2} &=&\frac{\alpha }{(n-2)(n-3)}, \\
\alpha _{3} &=&\frac{\alpha ^{2}}{72(_{\phantom{-}{4}}^{n-2})},
\end{eqnarray*}
where $m$ is the mass parameter. Solutions of Gauss-Bonnet gravity are not
real in the whole range $0\leq r<\infty $ and one needs a transformation to
make them real \cite{DH,DehBord}. But, here the metric function $f(r)$ is
real in the whole range $0\leq r<\infty $.

In order to consider the asymptotic behavior of the solution, we put $m=q=0$
where the metric function reduces to
\begin{equation}
f(r)=k+\frac{r^{2}}{\alpha }\left[ 1-\left( 1+\frac{6\Lambda \alpha }{n(n-1)}%
\right) ^{1/3}\right].  \label{Fg0}
\end{equation}
Equation (\ref{Fg0}) shows that the asymptotic behavior of the solution is
AdS or dS provided $\Lambda <0$ or $\Lambda >0$. The case of asymptotic flat
solutions ($\Lambda =0$) is permitted only for $k=1$.

As in the case of black holes of Gauss-Bonnet-Born-Infeld
gravity \cite{Wil2,DH}, the above metric given by Eqs. (\ref{met}), (\ref{fr})
and (\ref{gr}) has an essential timelike singularity at $r=0$. Seeking
possible black hole solutions, we turn to looking for the existence of
horizons. The event horizon(s), if there exists any, is (are) located at the
root(s) of $g^{rr}=f(r)=0$. Denoting the largest real root of $f(r)$ by $%
r_{+}$, we consider first the case that $f(r)$ has only one real root. In
this case $f(r)$ is minimum at $r_{+}$ and therefore $f^{\prime }(r_{+})=0$.
That is,
\begin{equation}
(n-1)k\left[ 3(n-2)r_{+}^{4}+3(n-4)k\alpha r_{+}^{2}+(n-6)k^{2}\alpha ^{2}%
\right] +12r_{+}^{6}\beta ^{2}\left( 1-\sqrt{1+\eta _{+}}\right) -6{\Lambda }%
r_{+}^{6}=0  \label{f'}.
\end{equation}
One can find the extremal value of mass, $m_{\mathrm{ext}}$, in terms of
parameters of metric function by finding $r_{+}$ from Eq. (\ref{f'}) and
inserting it into equation $f(r_{+})=0$. Then, the metric of Eqs. (\ref{met}%
), (\ref{fr}) and (\ref{gr}) presents a black hole solution with inner and
outer event horizons provided $m>m_{\mathrm{ext}}$, an extreme black hole
for $m=m_{\mathrm{ext}}$ [temperature is zero since it is proportional to $%
f^{\prime }(r_{+})$] and a naked singularity otherwise. It is a matter of
calculation to show that $m_{\mathrm{ext}}$ for $k=0$ becomes
\begin{equation}
m_{\mathrm{ext}}=\frac{2(n-1)q_{\mathrm{ext}}^{2}}{n}\left( \frac{\Lambda
(\Lambda -4\beta ^{2})}{2(n-1)(n-2)\beta ^{2}q_{\mathrm{ext}}^{2}}\right)
^{(n-2)/(2n-2)}\digamma (\frac{\Lambda (\Lambda -4\beta ^{2})}{4\beta ^{4}}).
\label{mext}
\end{equation}

The Hawking temperature of the black holes can be easily obtained by
requiring \ the absence of conical singularity at the horizon in the
Euclidean sector of the black hole solutions. One obtains
\begin{equation}
{T}_{+}=\frac{f^{\prime }(r_{+})}{4\pi }
=\frac{ (n-1)k\left[
3(n-2)r_{+}^{4}+3(n-4)k\alpha r_{+}^{2}+(n-6)k^{2}\alpha ^{2}\right]
+12r_{+}^{6}\beta ^{2}\left( 1-\sqrt{1+\eta _{+}}\right) -6{\Lambda }%
r_{+}^{6}} {12\pi (n-1)r_{+}(r_{+}^{2}+k\alpha )^{2}}.
\label{Temp}
\end{equation}
It is worthwhile to note that $T_{+}$ is zero for $m=m_{\mathrm{ext}}$.

\section{Thermodynamics of Asymptotically Flat Black Holes for $k=1$}

\label{flat}

In this section, we consider the thermodynamics of spherically symmetric
black holes which are asymptotically flat. This is due to the fact that only
the entropy of asymptotically black holes of Lovelock gravity is well known \cite{Myer}.
Usually entropy of black holes
satisfies the so-called area law of entropy which states that the black hole
entropy equals one-quarter of the horizon area \cite{Bek}. One of the
surprising and impressive features of this area law of entropy is its
universality. It applies to all kinds of black holes and black strings of
Einstein gravity \cite{Haw}. However, in higher derivative gravity the area
law of entropy is not satisfied in general \cite{fails}. It is known that
the entropy of asymptotically flat black holes of Lovelock gravity is \cite{Myer}
\begin{equation}
S=\frac{1}{4}\sum_{k=1}^{[(n-2)/2]}k\alpha _{k}\int d^{n-1}x\sqrt{\tilde{g}}%
\tilde{\mathcal{L}}_{k-1}, \label{Enta}
\end{equation}
where the integration is done on the $(n-1)$-dimensional spacelike
hypersurface of the Killing horizon, $\tilde{g}_{\mu \nu }$ is the induced
metric on it, $\tilde{g}$ is the determinant of $\tilde{g}_{\mu \nu }$, and $%
\tilde{\mathcal{L}}_{k}$ is the $k$th order Lovelock Lagrangian of $\tilde{g}%
_{\mu \nu }$. Thus, the entropy for asymptotically flat black holes in third order Lovelock gravity is
\begin{equation}
S=\frac{1}{4}\int d^{n-1}x\sqrt{\tilde{g}}\left( 1+2\alpha _{2}\tilde{R}%
+3\alpha _{3}(\tilde{R}_{\mu \nu \sigma \kappa }\tilde{R}^{\mu \nu \sigma
\kappa }-4\tilde{R}_{\mu \nu }\tilde{R}^{\mu \nu }+\tilde{R}^{2})\right),
\label{Entb}
\end{equation}
where $\tilde{R}_{\mu \nu \rho \sigma }$ and $\tilde{R}_{\mu \nu }$ are
Riemann and Ricci tensors and $\tilde{R}$ is the Ricci scalar for the
induced metric $\tilde{g}_{ab}$ on the $(n-1)$-dimensional horizon. It is a
matter of calculation to show that the entropy of black holes is
\begin{equation}
S=\frac{V_{n-1}}{4}\left( r_{+}^{4}+\frac{2(n-1)}{n-3}\alpha r_{+}^{2}+\frac{%
n-1}{n-5}\alpha ^{2}\right)r_{+}^{n-5}. \label{Ent1}
\end{equation}

The charge of the black hole can be found by calculating the flux of the
electric field at infinity, yielding
\begin{equation}
Q=\frac{V_{n-1}}{4\pi }\sqrt{\frac{(n-1)(n-2)}{2}}q . \label{Ch}
\end{equation}
The electric potential $\Phi $, measured at infinity with respect to the
horizon, is defined by
\begin{equation}
\Phi =A_{\mu }\chi ^{\mu }\left\vert _{r\rightarrow \infty }-A_{\mu }\chi
^{\mu }\right\vert _{r=r_{+}},  \label{Pot1}
\end{equation}
where $\chi =\partial /\partial t$ is the null generator of the horizon. One
finds
\begin{equation}
\Phi =\sqrt{\frac{(n-1)}{2(n-2)}}\frac{q}{r_{+}^{n-2}}\digamma (\eta _{+}).
\label{Pot2}
\end{equation}
The ADM (Arnowitt-Deser-Misner) mass of black hole can be obtained by using the behavior of the
metric at large $r$. It is easy to show that the mass of the black hole is
\begin{equation}
M=\frac{V_{n-1}}{16\pi }\left( n-1\right) m.  \label{Mass1}
\end{equation}

We now investigate the first law of thermodynamics. Using the expression for
the entropy, the charge, and the mass given in Eqs. (\ref{Ent1}), (\ref{Ch})
and (\ref{Mass1}), and the fact that $f(r_{+})=0$, one obtains

\begin{eqnarray}
M(S,Q) &=&\frac{\left( n-1\right) }{16\pi }\left\{ \frac{2r_{+}^{n}}{n(n-1)}%
\left( 2\beta ^{2}\left[ 1-\sqrt{1+\Im }+\frac{(n-1)\Im }{(n-2)}\digamma
(\Im )\right] -\Lambda \right) \right.  \nonumber \\
&&\left. -r_{+}^{n-2}+\alpha r_{+}^{n-4}-\frac{\alpha ^{2}r_{+}^{n-6}}{3}%
\right\},  \label{Sma}
\end{eqnarray}
where
\[
\Im =\frac{16\pi ^{2}Q^{2}}{\beta ^{2}r_{+}^{2n-2}}.
\]
In Eq. (\ref{Sma}) $r_{+}$\ is the real root of Eq. (\ref{Ent1}) which is a function of $S$%
. One may then regard the parameters $S$ and $Q$ as a complete set of
extensive parameters for the mass $M(S,Q)$ and define the intensive
parameters conjugate to them. These quantities are the temperature and the
electric potential
\begin{equation}
T=\left( \frac{\partial M}{\partial S}\right) _{Q},\ \ \ \ \Phi =\left(
\frac{\partial M}{\partial Q}\right) _{S}.  \label{Dsma1}
\end{equation}
Computing $\partial M/\partial r_{+}$ and $\partial S/\partial r_{+}$ and
using the chain rule, it is easy to show that the intensive quantities
calculated by Eq. (\ref{Dsma1}) coincide with Eqs. (\ref{Temp}) and (\ref
{Pot2}), respectively. Thus, the thermodynamic quantities calculated in Eqs. (%
\ref{Temp}) and (\ref{Pot2}) satisfy the first law of thermodynamics,
\begin{equation}
dM=TdS+\Phi dQ . \label{1stlaw}
\end{equation}

\section{Thermodynamics of Asymptotically AdS Rotating Black Branes with
Flat Horizon}

\label{rotating}

Now, we want to endow our spacetime solution (\ref{met}) for $k=0$\ with a
global rotation. These kinds of rotating solutions in Einstein gravity
have been introduced in \cite{Lem}. In order to add angular momentum to the spacetime, we
perform the following rotation boost in the $t-\phi _{i}$ planes
\begin{equation}
t\mapsto \Xi t-a_{i}\phi _{i},\hspace{0.5cm}\phi _{i}\mapsto \Xi \phi _{i}-%
\frac{a_{i}}{l^{2}}t  \label{Tr}
\end{equation}
for $i=1...[n/2]$, where $[x]$ is the integer part of $x$. The maximum
number of rotation parameters is due to the fact that the rotation group in $%
n+1$ dimensions is $SO(n)$ and therefore the number of independent rotation
parameters is $[n/2]$. Thus the metric of an asymptotically AdS rotating
solution with $p\leq \lbrack n/2]$ rotation parameters for flat horizon can
be written as
\begin{eqnarray}
ds^{2} &=&-f(r)\left( \Xi dt-{{\sum_{i=1}^{p}}}a_{i}d\phi _{i}\right) ^{2}+%
\frac{r^{2}}{l^{4}}{{\sum_{i=1}^{p}}}\left( a_{i}dt-\Xi l^{2}d\phi
_{i}\right) ^{2}  \nonumber \\
&&\ \text{ }+\frac{dr^{2}}{f(r)}-\frac{r^{2}}{l^{2}}{\sum_{i<j}^{p}}%
(a_{i}d\phi _{j}-a_{j}d\phi _{i})^{2}+r^{2}{{\sum_{i=p+1}^{n-1}}}d\phi _{i},
\label{met1}
\end{eqnarray}
where $\Xi =\sqrt{1+\sum_{i}^{k}a_{i}^{2}/l^{2}}$. Using Eq. (\ref{BIeq}),
one can show that the vector potential can be written as
\begin{equation}
A_{\mu }=-\sqrt{\frac{(n-1)}{2n-4}}\frac{q}{r^{n-2}}\digamma (\eta )\left(
\Xi \delta _{\mu }^{0}-\delta _{\mu }^{i}a_{i}\right) \text{(no sum on }i%
\text{)}. \label{Amu}
\end{equation}

One can obtain the temperature and angular momentum of the event horizon by
analytic continuation of the metric. One obtains
\begin{equation}
{T}_{+}{=}\frac{f^{\prime }(r_{+})}{4\pi \Xi }={{\frac{r_{+}}{2(n-1)\pi \Xi }%
}}\left( 2\beta ^{2}(1-\sqrt{1+\eta _{+}})-{\Lambda }\right),  \label{Tem}
\end{equation}
\begin{equation}
\Omega _{i}=\frac{a_{i}}{\Xi l^{2}},  \label{Om}
\end{equation}
where $\eta _{+}=\eta (r=r_{+})$. Next, we calculate the electric charge and
potential of the solutions. The electric charge per unit volume $V_{n-1}$
can be found by calculating the flux of the electric field at infinity,
yielding
\begin{equation}
Q=\frac{1}{4\pi }\sqrt{\frac{(n-1)(n-2)}{2}}\,\,\Xi q. \label{Charge}
\end{equation}
Using Eq. (\ref{Pot1}) and the fact that $\chi =\partial _{t}+{\sum_{i}^{k}}%
\Omega _{i}\partial _{\phi _{i}}$ is the null generator of the horizon, the
electric potential $\Phi $ is obtained as
\begin{equation}
\Phi =\sqrt{\frac{(n-1)}{2(n-2)}}\frac{q}{\Xi r_{+}^{n-2}}\digamma (\eta
_{+}).  \label{Pot}
\end{equation}

\subsection{Conserved quantities of the solutions}

Here, we calculate the action and conserved quantities of the black brane
solutions. In general the action and conserved quantities of the spacetime
are divergent when evaluated on the solutions. A systematic method of
dealing with this divergence for asymptotically AdS solutions of Einstein
gravity is through the use of the counterterms method inspired by the
anti-de Sitter conformal field theory (AdS/CFT) correspondence \cite{Mal}.
For asymptotically AdS solutions of Lovelock gravity with flat boundary, $%
\widehat{R}_{abcd}(\gamma )=0$, the finite action is \cite{DBS,DM1}
\begin{equation}
I=I_{G}+\frac{1}{8\pi }\int_{\partial \mathcal{M}}d^{n}x\sqrt{-\gamma }%
\left\{ L_{1b}+\alpha _{2}L_{2b}+\alpha _{3}L_{3b}\right\} +\frac{1}{8\pi }%
\int_{\partial \mathcal{M}}d^{n}x\sqrt{-\gamma }\left( \frac{n-1}{L}\right),
\label{Ifinite}
\end{equation}
where $L$\ is
\begin{eqnarray}
L &=&\frac{15l^{2}\sqrt{\alpha (1-\lambda )}}{5l^{2}+9\alpha -l^{2}\lambda
^{2}-4l^{2}\lambda }, \\
\lambda &=&(1-\frac{3\alpha }{l^{2}})^{1/3}.
\end{eqnarray}
One may note that $L$ reduces to $l$\ as $\alpha $\ goes to zero. The first integral in Eq. (%
\ref{Ifinite}) is a boundary term which is chosen such that the variational
principle is well defined. In this integral $L_{1b}=K$, $L_{2b}=2(J-2%
\widehat{G}_{ab}^{(1)}K^{ab})$ and
\[
L_{3b}=3(P-2\widehat{G}_{ab}^{(2)}K^{ab}-12\widehat{R}_{ab}J^{ab}+2\widehat{R%
}J-4K\widehat{R}_{abcd}K^{ac}K^{bd}-8\widehat{R}%
_{abcd}K^{ac}K_{e}^{b}K^{ed}),
\]
where $\gamma _{\mu \nu }$ and $K$ are induced metric and trace of extrinsic
curvature of boundary, $\widehat{G}_{ab}^{(1)}$\textbf{\ }and\textbf{\ }$%
\widehat{G}_{ab}^{(2)}$\textbf{\ }are the $n$-dimensional Einstein and
second order Lovelock tensors (Eq. (\ref{Love2})) of the metric $\gamma
_{ab} $\ and $J$\ and $P$\ are the trace of
\begin{equation}
J_{ab}=\frac{1}{3}%
(2KK_{ac}K_{b}^{c}+K_{cd}K^{cd}K_{ab}-2K_{ac}K^{cd}K_{db}-K^{2}K_{ab}),
\end{equation}
and
\begin{eqnarray}
P_{ab} &=&\frac{1}{5}%
\{[K^{4}-6K^{2}K^{cd}K_{cd}+8KK_{cd}K_{e}^{d}K^{ec}-6K_{cd}K^{de}K_{ef}K^{fc}+3(K_{cd}K^{cd})^{2}]K_{ab}
\nonumber \\
&&-(4K^{3}-12KK_{ed}K^{ed}+8K_{de}K_{f}^{e}K^{fd})K_{ac}K_{b}^{c}-24KK_{ac}K^{cd}K_{de}K_{b}^{e}
\nonumber \\
&&+(12K^{2}-12K_{ef}K^{ef})K_{ac}K^{cd}K_{db}+24K_{ac}K^{cd}K_{de}K^{ef}K_{bf}\}
\label{Pab}.
\end{eqnarray}
Using Eqs. (\ref{Act1}) and (\ref{Ifinite}), the finite action per unit
volume $V_{n-1}$ can be calculated as
\begin{equation}
I=-\frac{1}{T_{+}}\left\{ \frac{r_{+}^{n}}{16\pi l^{2}}-\frac{r_{+}^{n}\beta
^{2}(\sqrt{1+\eta _{+}}-1)}{4n(n-1)\pi }+\frac{(n-1)q^{2}}{8n\pi
r_{+}^{(n-2)}}\digamma (\eta _{+})\right\}.  \label{finiteAct}
\end{equation}
Using the Brown-York method \cite{Brown}, the finite energy-momentum tensor
is
\begin{equation}
T^{ab}=\frac{1}{8\pi }\{(K^{ab}-K\gamma ^{ab})+2\alpha _{2}(3J^{ab}-J\gamma
^{ab})+3\alpha _{3}(5P^{ab}-P\gamma ^{ab})+\frac{n-1}{L}\gamma ^{ab}\ \},
\end{equation}
and the conserved quantities associated with the Killing vectors $\partial
/\partial t$ and $\partial /\partial \phi ^{i}$ are
\begin{eqnarray}
M &=&\frac{1}{16\pi }m\left( n\Xi ^{2}-1\right) ,  \label{Mass} \\
J_{i} &=&\frac{1}{16\pi }n\Xi ma_{i},  \label{Angmom}
\end{eqnarray}
which are the mass and angular momentum of the solution.

Now using Gibbs-Duhem relation
\begin{equation}
S=\frac{1}{T}(M-Q\Phi -{{\sum_{i=1}^{k}}}\Omega _{i}J_{i})-I,
\label{GibsDuh}
\end{equation}
and Eqs. (\ref{Pot}), (\ref{finiteAct}) and (\ref{Mass})-(\ref{Angmom}) one obtains
\begin{equation}
S=\frac{\Xi }{4}r_{+}^{n-1}  \label{Entropy}
\end{equation}
for the entropy per unit volume $V_{n-1}$. This shows that the entropy obeys
the area law for our case where the horizon is flat.

\subsection{Stability of the solutions\label{Stab}}

Calculating all the thermodynamic and conserved quantities of the black
brane solutions, we now check the first law of thermodynamics for our
solutions with flat horizon. We obtain the mass as a function of the
extensive quantities $S$, $\mathbf{J}$, and $Q$. Using the expression for charge
mass, angular momenta and entropy given in Eqs. (\ref{Charge}),
(\ref{Mass}), (\ref{Angmom}), (\ref{Entropy}) and the fact that $%
f(r_{+})=0$, one can obtain a Smarr-type formula as
\begin{equation}
M(S,\mathbf{J},Q)=\frac{(nZ-1)J}{nl\sqrt{Z(Z-1)}},  \label{Smar}
\end{equation}
where $J=\left\vert \mathbf{J}\right\vert =\sqrt{\sum_{i}^{k}J_{i}^{2}}$ and
$Z=\Xi ^{2}$ is the positive real root of the following equation:
\begin{equation}
\frac{Z^{1/2(n-1)}}{\sqrt{Z-1}}=\frac{\left[ 4l^{2}\beta ^{2}\left( 1-\sqrt{%
1+\eta _{+}}\right) +n(n-1)\right] S^{n/(n-1)}}{(n-1)\pi lJ}+\frac{4\pi
lQ^{2}\digamma (\frac{\pi ^{2}Q^{2}}{\beta ^{2}S^{2}})}{%
(n-2)J(4S)^{(n-2)/(n-1)}}.
\end{equation}
One may then regard the parameters $S$, $J_{i}$'s, and $Q$ as a complete set
of extensive parameters for the mass $M(S,\mathbf{J},Q)$ and define the
intensive parameters conjugate to them. These quantities are the
temperature, the angular velocities, and the electric potential
\begin{equation}
T=\left( \frac{\partial M}{\partial S}\right) _{J,Q},\ \ \Omega _{i}=\left(
\frac{\partial M}{\partial J_{i}}\right) _{S,Q},\ \ \Phi =\left( \frac{%
\partial M}{\partial Q}\right) _{S,J}.  \label{Dsmar}
\end{equation}
Straightforward calculations show that the intensive quantities calculated
by Eq. (\ref{Dsmar}) coincide with Eqs. (\ref{Tem}), (\ref{Om}) and (\ref
{Pot}). Thus, these quantities satisfy the first law of thermodynamics:
\[
dM=TdS+{{{\sum_{i=1}^{k}}}}\Omega _{i}dJ_{i}+\Phi dQ.
\]

Finally, we investigate the local stability of charged rotating black brane
solutions of third order Lovelock gravity in the presence of nonlinear
electrodynamic Born-Infeld field in the canonical and grand canonical
ensembles. In the canonical ensemble, the positivity of the heat capacity $%
C_{\mathbf{J},Q}=T_{+}/(\partial ^{2}M/\partial S^{2})_{\mathbf{J},Q}$ and
therefore the positivity of $(\partial ^{2}M/\partial S^{2})_{\mathbf{J},Q}$
is sufficient to ensure the local stability. Using the fact that
\[
{_{2}F_{1}}\left( \left[ {\frac{3}{2},\frac{3{n-4}}{{2n-2}}}\right] ,\left[ {%
\frac{5{n-6}}{{2n-2}}}\right] ,-z\right) =\frac{(3n-4)}{(n-1)z}\left\{
\digamma (\frac{\pi ^{2}Q^{2}}{\beta ^{2}S^{2}})-\frac{1}{\sqrt{1+z}}%
\right\},
\]
it is easy to show that
\begin{eqnarray}
&&\frac{\partial ^{2}M}{\partial S^{2}}=\frac{2[(n-1)(n-2)^{2}q^{2}-\Lambda
r_{+}^{(2n-2)}\sqrt{1+\eta _{+}}+2\beta ^{2}r_{+}^{(2n-2)}\left( \sqrt{%
1+\eta _{+}}-1\right) ]}{(n-1)^{2}\pi \Xi ^{2}r_{+}^{(3n-4)}\sqrt{1+\eta _{+}%
}}-  \nonumber \\
&&\frac{8(\Xi ^{2}-1)\Lambda \left( -r_{+}^{2n-2}\sqrt{1+\eta _{+}}(\frac{%
n(n-1)}{2}+2\beta ^{2}l^{2})+(n-1)(n-2)\beta l^{2}q^{2}+2\beta
^{3}l^{2}r_{+}^{(2n-2)}\right) ^{2}}{\pi ml^{2}\beta ^{2}\Xi
^{2}(n-1)^{4}(4\Xi ^{2}+1)(1+\eta _{+})r^{(4n-6)}}.  \label{dMSS}
\end{eqnarray}
Both of the two terms of Eq. (\ref{dMSS}) are positive, and therefore the
condition for thermal equilibrium in the canonical ensemble is satisfied.

In the grand canonical ensemble, the positivity of the determinant of the Hessian
matrix of $M(S,Q,\mathbf{J})$ with respect to its extensive variables $X_{i}$%
, $\mathbf{H}_{X_{i}X_{j}}^{M}=\left( \partial ^{2}M/\partial X_{i}\partial
X_{j}\right) $, is sufficient to ensure the local stability. It is a matter
of calculation to show that the determinant of $\mathbf{H}_{S,Q,\mathbf{J}%
}^{M}$ is:
\begin{equation}
\left\vert \mathbf{H}_{SJQ}^{M}\right\vert =\frac{64\pi \left( 2(n-2)\beta
^{2}\eta _{+}-\Lambda \sqrt{1+\eta _{+}}+2\beta ^{2}\left( \sqrt{1+\eta _{+}}%
-1\right) \right) }{(n-2)(n-1)^{3}ml^{2}\Xi ^{6}r_{+}^{2(n-2)}\left[
(n-2)\Xi ^{2}+1\right] \sqrt{1+\eta _{+}}}\digamma ({\eta }_{+})+\frac{%
8(n-1)\Xi }{r_{+}}T_{+}.  \label{Hes}
\end{equation}
Equation (\ref{Hes}) shows that the determinant of the Hessian matrix is
positive, and therefore the solution is stable in the grand canonical ensemble
too. The stability analysis given here shows that the higher curvature
and nonlinear Maxwell terms in the action have no effect on the stability of black holes with flat horizon,
and these kinds of black holes are thermodynamically stable as in the case of toroidal black holes of
Einstein-Maxwell gravity \cite{Lem2}. This phase behavior is also commensurate with the fact that there is no
Hawking-Page transition for a black object whose horizon is diffeomorphic to
$\Bbb{R}^{p}$ and therefore the system is always in the high temperature
phase \cite{Wit2}.

\section{ CLOSING REMARKS}
In this paper we considered both the nonlinear scalar terms constructing
from the curvature tensor and electromagnetic field tensor in gravitational
action, which are on similar footing with regard to the string
corrections on gravity and electrodynamic sides. We presented static
topological black hole solutions of third order Lovelock gravity in the
presence of Born-Infeld gravity, which are asymptotically AdS for negative
cosmological constant, dS for positive $\Lambda $. For the case of solutions
with positive curvature horizon ($k=1$), one can also have asymptotically
flat solutions provided $\Lambda =0$. The topological solutions obtained in
this paper may be interpreted as black holes with two inner and outer event
horizons for $m>m_{\mathrm{ext}}$ , extreme black holes for $m=m_{\mathrm{ext%
}}$ or naked singularity otherwise. We found that these solutions reduce to
the solutions of Einstein-Born-Infeld gravity as the Lovelock coefficients
vanish, and reduce to the solutions of third order Lovelock gravity in the
presence of Maxwell field as $\beta $ goes to infinity \cite{DSh}. We
consider thermodynamics of asymptotically flat solutions and found that the
first law of thermodynamics is satisfied by the conserved and thermodynamic
quantities of the black hole. We also consider the rotating solution with
flat horizon and computed the action and conserved quantities of it through
the use of counterterm method. We found that the entropy obeys the area law
for black branes with flat horizon. We obtained a Smarr-type formula for the
mass of the black brane as a function of the entropy, the charge, and the
angular momenta, and found that the conserved and thermodynamics quantities
satisfy the first law of thermodynamics. We also studied the phase behavior
of the $(n+1)$-dimensional rotating black branes in third order Lovelock
gravity and showed that there is no Hawking-Page phase transition in spite
of the angular momenta of the branes and the presence of a nonlinear
electromagnetic field. Indeed, we calculated the heat capacity and the
determinant of the Hessian matrix of the mass with respect to $S$, $\mathbf{J%
}$, and $Q$ of the black branes and found that they are positive for all the
phase space, which means that the brane is locally stable for all the
allowed values of the metric parameters.
\acknowledgments{This work has been supported by Research
Institute for Astrophysics and Astronomy of Maragha.}

\end{document}